# Lattice effects on ferromagnetism in perovskite ruthenates

J.-G. Cheng, J.-S. Zhou*, and J.B. Goodenough†

The Materials Science and Engineering Program/ Mechanical Engineering, University of Texas at Austin, Austin, Texas 78712, USA

**Ferromagnetism and its evolution in the orthorhombic perovskite system $Sr_{1-x}Ca_xRuO_3$ have been widely believed to correlate with the structural distortion. The recent development of high-pressure synthesis of the Ba substituted $Sr_{1-y}Ba_yRuO_3$ makes it possible to study ferromagnetism over a broader phase diagram, which includes new orthorhombic *Imma* and the cubic phases. However, the chemical substitutions introduce the A-site disorder effect on $T_c$, which complicates determination of the relationship between ferromagnetism and the structural distortion. By clarifying the site disorder effect on $T_c$ in several new series of ruthenates in which the average bond length $\langle A\text{-}O\rangle$ keeps the same, but the bond-length variance varies, we are able to demonstrate a parabolic curve of $T_c$ versus mean bond length $\langle A\text{-}O\rangle$. A much higher $T_c \approx 177$ K than that found in orthorhombic $SrRuO_3$ can be obtained from the curve at a bond length $\langle A\text{-}O\rangle$ which makes the geometric factor $t = \langle A\text{-}O\rangle/(\sqrt{2}\langle Ru\text{-}O\rangle) \approx 1$. This new result reveals not only that the ferromagnetism in the ruthenates is extremely sensitive to the lattice strain, but also that it has an important implication for exploring the structure-property relationship in a broad range of oxides with perovskite or perovskite-related structure.**

The A-site cation in conductive transition-metal oxides $AMO_3$ with a perovskite or a perovskite-related structure does not contribute directly to the density of electronic states near the Fermi energy. However, both the superconductive transition temperature $T_c$ in the cuprates (1, 2) and the ferromagnetic transition temperature $T_c$ in the magnetoresistive manganites (3) are highly sensitive to the A-cation disorder as well as the mean A-cation radius $\langle r_A\rangle$ in these mixed-valent systems. The lattice effects are greatly enhanced in manganites where the ferromagnetic transition is accompanied by a first-order metal-insulator transition. Ferromagnetism in the orthorhombic perovskite $SrRuO_3$ and its unusual disappearance in isostructural $CaRuO_3$ (4-8) have not been well-understood. Early band structure calculations indicated that the orthorhombic distortion is critical for the electronic structure that supports ferromagnetism in both $SrRuO_3$ and $CaRuO_3$.(9) Mazin and Singh (10) have argued from their first-principles calculation that the orthorhombic distortion in $CaRuO_3$ is responsible for a much reduced $N(E_F)$ and the suppression of ferromagnetism. On the other hand, they have found an unusually high $N(E_F)$ in the cubic phase that can lead to instabilities. The small orthorhombic distortion found in $SrRuO_3$ was considered nearly ideal to stabilize ferromagnetism. In a recent calculation, Middey *et al* (11) have further argued that an optimum $T_c$ in the perovskite ruthenates would occur in the orthorhombic phase where the $GdFeO_3$-type distortion is small. An enlarged orthorhombic distortion and smaller cell volume was found to destroy ferromagnetism in $CaRuO_3$. These models have been used to explain qualitatively the experimental phase diagram of $T_c$ versus $\langle r_A \rangle$ for $Sr_{1-x}Ca_xRuO_3$ available in the literature. Extending the phase diagram to the Ba-substituted side brings a new test to our existing understanding. Recent progress in the high-pressure synthesis of $Sr_{1-y}Ba_yRuO_3$ has demonstrated the entire evolution of the ferromagnetic phase all the way to $BaRuO_3$.(12) Unlike the rare-earth substitution in $RMO_3$ perovskite systems, the change from Ca to Sr and to Ba introduces a relatively large change of $\langle r_A \rangle$ in addition to a slightly different chemical property between them. Combinations between these three elements make possible a continuous change of $\langle r_A \rangle$; but it also introduces the size variance effect on $T_c$. It is therefore not clear whether $T_c$ truly peaks out at the $\langle r_A \rangle$ for $SrRuO_3$ as shown by an experimental phase diagram of $A_{1-x}A'_xRuO_3$ (A,A' =Ca, Sr, Ba) (12) and as expected by first-principles calculations. In order to obtain how $T_c$ is correlated to the crystal structure, which is essential for understanding ferromagnetism in ruthenates, we have made a systematic study of the size variance effect on $T_c$. This study leads to a new phase diagram of $T_c$ versus the average $\langle A-O \rangle$ bond length without influence from the bond-length variance. These new results indicate that the current understanding about ferromagnetism in ruthenates is incomplete.

We used the following procedure to introduce the A-cation size variance. The tabulated ionic radii depend sensitively on the coordination number (13) and they have been found to be not compatible with structures determined by the state-of-the-art neutron diffractions in recent years. Moreover, the $O^{2-}$ ion should not be treated as a hard sphere. We have chosen the bond lengths obtained experimentally on perovskite ruthenates as listed in Table I.(12, 14, 15) One may be concerned that the bond length ⟨Ba-O⟩ = 2.832 Å from a cubic $BaRuO_3$ is under compression whereas the ⟨Ru-O⟩ = 2.003 Å is under tension (12); therefore, some adjustment may be needed in order to have the equilibrium mean bond length for calculating the ⟨A-O⟩ for other compositions with the Ba substitution. First of all, we have no basis for the adjustment. Second, it may also not be necessary since the substituted Ba would create the same local lattice strain as in a cubic $BaRuO_3$. As a matter of fact, all compositions having a Ba component need to be synthesized under pressure regardless of their ⟨A-O⟩ relative to ⟨Sr-O⟩, which may justify that we have used ⟨Ba-O⟩ = 2.832 Å from a high-pressure product $BaRuO_3$ for calculating the ⟨A-O⟩ for other compositions. We have obtained essentially the same conclusion by using the ⟨A-O⟩ bond length either from experiment or the tabulated value. In order to have all samples made under the same conditions, we have also made $SrRuO_3$ by high-pressure synthesis. $T_c$ of the high-pressure-synthesized (HP) $SrRuO_3$ increases by 4 K and the moment at 5 K reaches 1.6 $\mu_B$ under 5 T, which is equivalent to its value at 44 T for a sample prepared at ambient pressure.(16) Based on the bond lengths in Table I, a combination of 0.51⟨Ca-O⟩ and 0.49 ⟨Ba-O⟩ is equal to ⟨Sr-O⟩. Therefore, the substitution of $Ba_{.49}Ca_{.51}$ for Sr in $Sr_{1-x}(Ba_{.49}Ca_{.51})_xRuO_3$ does not change the mean A-O bond length ⟨A-O⟩ = ⟨Sr-O⟩, but it changes the bond-length variance defined as $\sigma^2 = \sum a_i L_i^2 - (\sum a_i L_i)^2$. The Sr concentration in $Sr_{1-x}Ca_xRuO_3$ and $Sr_{1-y}Ba_yRuO_3$ has been progressively replaced by $Ba_{.49}Ca_{.51}$; therefore the bond-length variance can be varied in a series of ruthenates for a given constant ⟨A-O⟩. The pressure required for synthesizing a single phase sample increases with increasing overall Ba concentration. At P =15 GPa, the highest pressure in this study, the maximum concentration of ($Ba_{.49}Ca_{.51}$) substitution for Sr in $Sr_{1-x}Ca_xRuO_3$ is 0.35 for $SrRuO_3$; its substitution for Sr in $Sr_{1-y}Ba_yRuO_3$ progressively decreases.

Fig.1 shows the temperature dependence of magnetic susceptibility χ for some typical ruthenates. While the paramagnetic $\chi^{-1}(T)$ of $SrRuO_3$ and the Ba substituted $Sr_{1-y}Ba_yRuO_3$ fulfills the typical Curie-Weiss law, an anomaly in $\chi^{-1}(T)$, pointed by an arrow, can be found in the paramagnetic phase of $Sr_{1-x}Ca_xRuO_3$ at $T > T_c$. A similar anomaly has also been observed by He *et al* (6) for members of the $Sr_{1-x}Ca_xRuO_3$ system. Although $T_c$ is reduced as the bond-length variance is introduced, for example, in the series $Sr_{1-x}(Ba_{.49}Ca_{.51})_xRuO_3$ with ⟨A-O⟩ = ⟨Sr-O⟩, the paramagnetic χ(T) can be well-described by the Curie-Weiss law. $T_c$ in these samples is defined to be where dχ(T)/dT shows a peak.

We start the discussion of the bond-length variance effect from the series of $Sr_{1-x}(Ba_{.49}Ca_{.51})_xRuO_3$ with ⟨A-O⟩ = ⟨Sr-O⟩ =2.78 Å; the bond-length variance $\sigma^2$ increases with x. As shown in Fig.2, $T_c$ decreases linearly with increasing $\sigma^2$ as described by the formula $T_c = T_c^0 - p\sigma^2$, where $T_c^0$ is the Curie temperature of the sample without the bond-length variance. The local disorder due to the bond-length variance introduces perturbations that are normally harmful for a long-range ordered state. Interestingly, the same bond-length variance effect on transition temperature $T_c$ has been found in both the high-$T_c$ cuprates and the ferromagnetic manganites $R_{1-x}A_xMnO_3$ where R is a rare-earth, A is an alkaline earth (after the size variance is converted into the bond-length variance) (1, 3) We have also synthesized a series of ruthenates with ⟨A-O⟩ ≠ ⟨Sr-O⟩ and having different $\sigma^2$. The same linear behavior of $T_c$ versus $\sigma^2$ has also been found in these series of samples. The difference is that the $T_c^0$ from the linear fitting can be compared directly with experiment only for the series with ⟨A-O⟩ = ⟨Sr-O⟩ =2.78 Å. For a series with ⟨A-O⟩ ≠ ⟨Sr-O⟩, $T_c^0$ can be obtained by extrapolating lines to $\sigma^2 = 0$. It should be noticed that lines fitting $T_c$ versus $\sigma^2$ for the whole series are nearly parallel to each other. Therefore, the averaged slope *p* from lines with multiple points can be used to predict $T_c^0$ for the series where only one or two points of ($T_c$, $\sigma^2$) are available. These results allow us to map out $T_c^0$ versus ⟨A-O⟩ for the entire family of perovskites $ARuO_3$ in Fig.3(a). It is clear that $T_c^0 = T_c$ for $SrRuO_3$ is no longer located at a maximum in the new phase diagram. The overall evolution of $T_c^0$ versus ⟨A-O⟩ can be fit nearly perfectly by a quadratic formula $T_c^0 = T_c^m - q(L^m - \langle A-O \rangle)^2$. The curve fitting gives a $T_c^0 = 0$ at $\langle A-O \rangle \approx$ 2.74 Å where the $T_c$ onset has been found experimentally. On the other side of the dome, it is difficult to verify the point of $T_c^0 = 0$ at $\langle A-O \rangle \approx 2.842$ Å since an element like Ra with $\langle Ra-O \rangle > \langle Ba-O \rangle$ is required. The maximum $T_c^0 \approx 177$ K, much higher than $T_c \approx 164$ K for SrRuO$_3$, occurs at $\langle A-O \rangle \approx 2.791$ Å. Fitting to a quadratic formula indicates that the lattice strain plays a role in the ferromagnetism of the perovskite ruthenates. In order to understand why $T_c$ peaks out at an $\langle A-O \rangle > \langle Sr-O \rangle$ and why a lattice strain formula fits well the curve of $T_c$ versus $\langle A-O \rangle$, we turn to the results of our structural study.

In contrast to the Ca substituted samples Sr$_{1-x}$Ca$_x$RuO$_3$, which can be synthesized under ambient pressure, the crystal structures of the high-pressure products Sr$_{1-y}$Ba$_y$RuO$_3$ have not been refined systematically until this work. Our new results, shown in Fig.3(b), complete the structural phase diagram of the perovskites ARuO$_3$; all detailed structural parameters from refinements and typical results can be found in the Supplement Materials. Due to the well-known octahedral-site distortion,(17) the lattice parameter *b* crosses *a* in the orthorhombic *Pbnm* phase of Sr$_{1-x}$Ca$_x$RuO$_3$. The lattice parameter *b* > *a* holds for the *Pbnm* phase with rigid octahedra as predicted by using the software SPuDS; the prediction is also shown as the dashed lines in the Fig.3(b). The lattice parameter crossing also serves as a precursor for the phase transition to the *Imma* phase, which is commonly found in other A$^{2+}$B$^{4+}$O$_3$ perovskite families between the orthorhombic *Pbnm* phase and the cubic *Pm-3m* phase.(18, 19) It is clear from comparison of Fig.3 (a) and (b) that the orthorhombic *Imma* phase offers the highest $T_c$. The continuous change of lattice parameters versus $\langle A-O \rangle$ on crossing from the *Pbnm* to *Imma* and finally to the *Pm-3m* phase offers no useful information to explain why ferromagnetism prefers the *Imma* phase. By refining XRD patterns, we have further resolved the dependences of the bond length Ru-O and the bond angle Ru-O-Ru versus $\langle A-O \rangle$ in Fig.3(c). The bond-length mismatch in the *Pbnm* phase due to the geometric tolerance factor t ≡ (A-O)/√2(B-O) < 1 places the Ru-O bond under compression, which directly leads to the cooperative octahedral-site rotation and therefore to reduction of the Ru-O-Ru bond angle from 180°. On the other hand, the Ru-O bond is stretched in the cubic phase as $\langle A-O \rangle$ increases with the Ba substitution in $Sr_{1-y}Ba_yRuO_3$. The overall variation of the Ru-O bond length as a function of ⟨A-O⟩ is within ±0.01 Å, which actually reaches the resolution limit for the refinement of XRD patterns in this study. However, the Ru-O-Ru bond angle just reaches 180° at the *Imma/ Pm-3m* phase boundary where the Ru-O bond is under neither a compressive nor a stretching stress. By taking the ⟨A-O⟩$_m$ =2.795 Å from the parabolic curve fitting and the experimental ⟨Ru-O⟩$_m$ ≈ 1.980 Å in Fig.3, we found that the maximum $T_c^0$ occurs at t = 0.998. It is interesting to notice that the Ru-O bond length undergoes a minimum at t ≈ 1. A similar observation has been made in the (Ca,Sr,Ba)SnO$_3$ system. (18) Instead of a cubic structure, the orthorhombic *Imma* structure corresponding to t ≈ 1 must be related to the bond-length variance for the composition. Ferromagnetism in the ruthenates appears to prefer the phase with a nearly equilibrium Ru-O bond length. The bond-length analysis is highly consistent with the fit of $T_c$ versus ⟨A-O⟩ to a quadratic formula, indicating a lattice strain effect, which is missing in the first-principles calculations. The calculation based on an itinerant-electron model predicted a Fermi surface instability as the cubic phase is approached.(10, 11) Therefore, a maximum $T_c$ should occur in a slightly distorted *Pbnm* phase. This argument is not supported by our experimental results.

As shown in Fig.1, the paramagnetic susceptibility and its behavior in the vicinity of $T_c$ in the Ca-substituted $Sr_{1-x}Ca_xRuO_3$ is remarkably different from that of the Ba substituted $Sr_{1-y}Ba_yRuO_3$, which exhibits a typical Curie-Weiss behavior at all T > $T_c$. The reduction of $\chi^{-1}(T)$ of $Sr_{1-x}Ca_xRuO_3$ as $T_c$ is approached in the paramagnetic phase is stunningly similar to that of a Griffiths phase.(20, 21) This observation has led to a scenario that the $T_c$ reduction as x increases in $Sr_{1-x}Ca_xRuO_3$ is due to a dilution effect by non-magnetic ions. Given that SrRuO$_3$ showing the Curie-Weiss $\chi^{-1}(T)$ falls on the same side of the dome as the Ca substituted samples from our new results in Fig.3(a), the Griffiths dilution story is not sufficient to explain the $T_c$ reduction at ⟨A-O⟩ < 2.795 Å. The covalent competition between the A-O bond and the B-O bond in the perovskite structure could be another factor to influence Curie temperature. In the perovskites ARuO$_3$, the A-O σ bond competes for the O-2p electron that π bonds with the Ru. On the Ba-substituted side, the more basic Ba-O bond competes less strongly for the common O-2p electron, which strengthens the Ru-O π bonding and can broaden the π* band despite a stretching of the Ru-O bond length. $T_c$ decreases with increasing bandwidth in an itinerant-electron ferromagnet. While these effects may influence $T_c$ in higher orders, the model fails to account for a maximum $T_c$ occurring in the *Imma* phase and for an overall profile of $T_c$ versus ⟨A-O⟩.

While the lattice strain story explains the overall observations of $T_c$ versus ⟨A-O⟩ and the Ru-O bond stress due to the bonding mismatch in the perovskite structure, it is still difficult to account for the pressure dependence of $T_c$ in cubic $BaRuO_3$.(22) The Ru-O bond in $BaRuO_3$ is highly stretched and $T_c$ is reduced to about 60 K. $T_c$ would increase under high pressure based on the lattice strain story since pressure reduces the Ru-O bond length without bending the Ru-O-Ru bond angle from 180°. However, we have found that $T_c$ in $BaRuO_3$ decreases dramatically under pressure and $BaRuO_3$ becomes a paramagnetic metal down to 2 K under P = 8 GPa. This discrepancy reflects that a higher compressibility of the ⟨Ba-O⟩ bond than the ⟨Ru-O⟩ bond, which increases the Ru-O covalent bonding to broaden the $RuO_2$ π* band. A similar discrepancy has been observed in high-$T_c$ cuprates. In the plot of $T_c$ versus ⟨$r_A$⟩ for the cuprates with a single $CuO_2$ layer, $HgBa_2CuO_4$ is located at the maximum on the parabolic curve.(2) A $dT_c/dP < 0$ can be predicted from this plot. However, experiment has shown that $T_c$ increases under pressure.(23)

From symmetry considerations in a band model,(10) the highest density of states $N(E_F)$ can be obtained in the cubic phase due to the high state degeneracy. On the other hand, the data for the Ru-O bond length and the Ru-O-Ru bond angle ϕ in Fig.3(c) show an abrupt increase in ⟨Ru-O⟩ on entering the *Pbnm* phase where a cooperative rotation of the RuO6 octahedra reduces the Ru-O-Ru bond angle from 180°. The increase in ⟨Ru-O⟩ would appear to relax the elastic strain on the bond, but the equilibrium ⟨Ru-O⟩ bond length would be increased by the introduction of localized 4d electrons, which have a larger volume. The ferromagnetism of the Ru-4d electrons signals that the π* bands are narrow and approach the crossover to the localized-electron state where a large spin-orbit coupling creates a localized-electron state with a reduced total magnetic moment below the Griffiths anomaly. The jump in the ⟨Ru-O⟩ bond length in the *Pbnm* phase together with the appearance of the Griffiths phase anomaly in the susceptibility is consistent with the coexistence below the Griffiths temperature of localized-electron clusters that grow in volume with increasing x in the $Sr_{1-x}Ca_xRuO_3$ system as the crossover to localized-electron behavior is approached from the itinerant electron side. The transition from a narrow itinerant-electron band to a localized-electron phase is commonly first-order and characterized by a coexistence two phases if not by stabilization of a charge-density wave. The existence of ferromagnetic order in the itinerant-electron phase can be expected to suppress formation of a spin-paired charge-density wave. The dilution in spin-spin interactions is caused by localized-electron clusters with an enhanced spin-orbit coupling that alters the spin-spin interactions responsible for the ferromagnetic itinerant-electron phase. This observation and the lattice strain effect on $T_c$ must be taken into account in modeling the mysterious ferromagnetism in perovskite ruthenates.

In conclusion, the availability of $Sr_{1-y}Ba_yRuO_3$ synthesized under high pressure extends the ferromagnetic phase into the orthorhombic *Imma* and cubic *Pm-3m* phases. We have shown that $T_c$ decreases linearly with the bond-length variance $\sigma^2$ in both $Sr_{1-y}Ba_yRuO_3$ and $Sr_{1-x}Ca_xRuO_3$. A ferromagnetic transition temperature $T_c^0$ without the influence from the bond-length variance can be obtained for $SrRuO_3$ and $BaRuO_3$ and for ⟨$r_A$⟩ other than A= Sr and Ba by extrapolating $T_c$ versus $\sigma^2$ to $\sigma^2=0$. The curve $T_c^0$ versus mean bond length ⟨A-O⟩ can be fit with a quadratic formula, which indicates that a lattice strain effect plays a dominant role in controlling the ferromagnetic transition of perovskite ruthenates. A maximum $T_c$ can be obtained in a perovskite ruthenate with the geometric tolerance factor t = 1 where the unstrained Ru-O bond length with 180° Ru-O-Ru bond angle is found.

**Materials and Methods**

The high-pressure syntheses were carried out in a Walker-type multianvil module (Rockland Research). All samples were examined with powder X-ray diffraction (XRD) at room temperature with a Philips X'pert diffractometer (Cu Kα radiation). Lattice parameters were obtained by using the least-square analysis with the software *Jade* and the Rietveld refinement method with the software *Fullprof*. Magnetic properties were measured with a superconducting quantum interference device (SQUID) magnetometer (Quantum Design).

**Acknowledgements**

The work was supported by NSF (DMR 1122603) and the Robert A. Welch foundation (Grant F-1066).

* jszhou@mail.utexas.edu

† [jgooenough@mail.utexas.edu](jgooenough@mail.utexas.edu)

Author contributions: all authors designed the research and wrote the paper; JGC and JSZ performed the materials synthesis, characterization, and analyzed the data.

Table I Experimental <A-O> bond lengths obtained on perovskite ruthenates $ARuO_3$

| $ARuO_3$ | <A-O> (Å) | <Ru-O> (Å) | References |
|---|---|---|---|
| A = Ca | 2.728 | 1.992 | 14 |
| A = Sr | 2.780 | 1.984 | 15 |
| A = Ba | 2.832 | 2.003 | 12 |

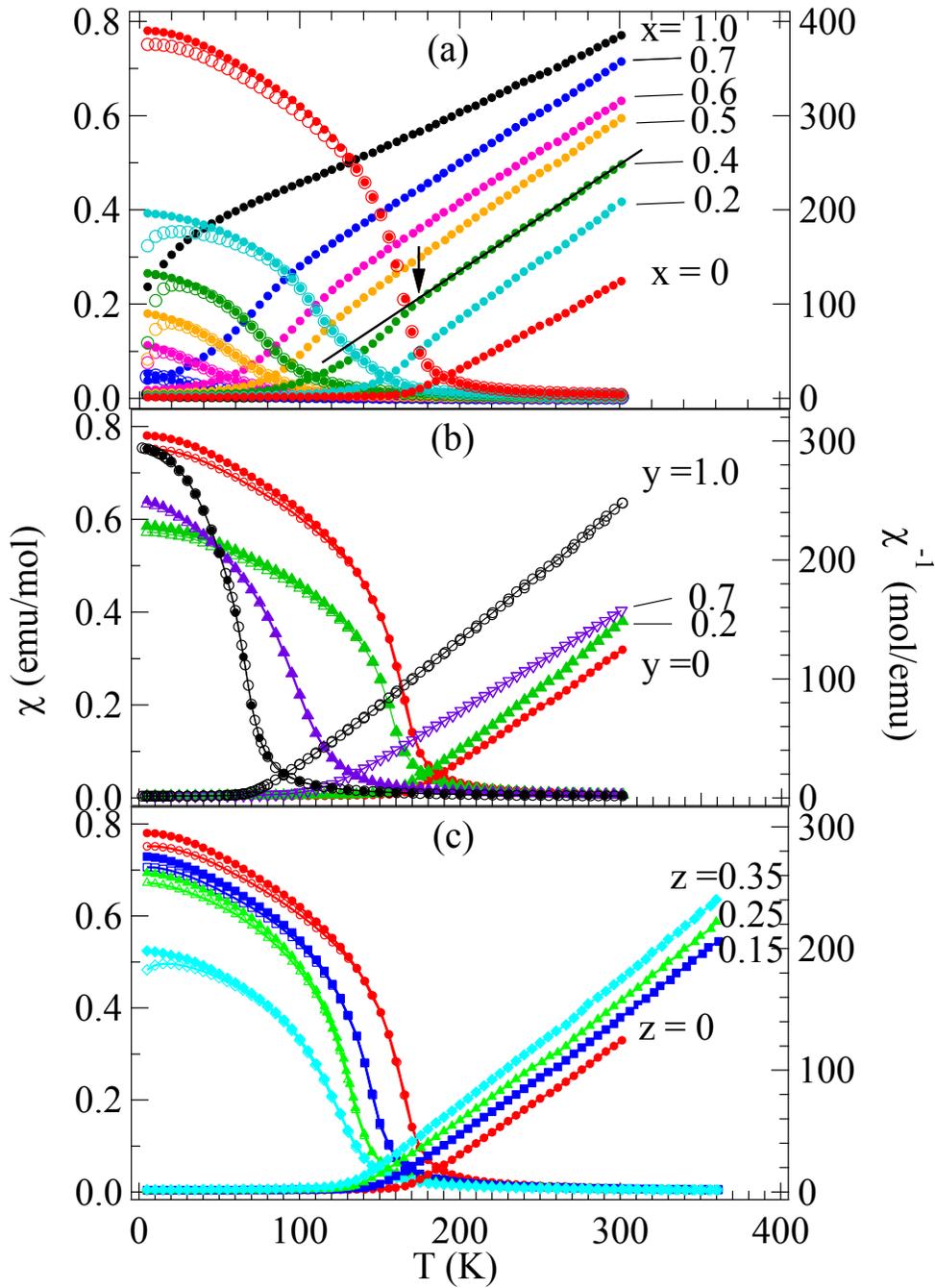

Fig. 1 Temperature dependences of magnetic susceptibility $\chi$ and $\chi^{-1}$ for (a) $Sr_{1-x}Ca_xRuO_3$; (b) $Sr_{1-y}Ba_yRuO_3$; (c) $Sr_{1-z}(Ba_{0.49}Ca_{0.51})_zRuO_3$. All measurements were performed with H = 1 T; open symbols are for zero-field-cool and solid symbols are for field cool. The arrow in (a) indicates the temperature where the $\chi^{-1}(T)$ deviates from a Curie-Weiss fitting in the paramagnetic phase.

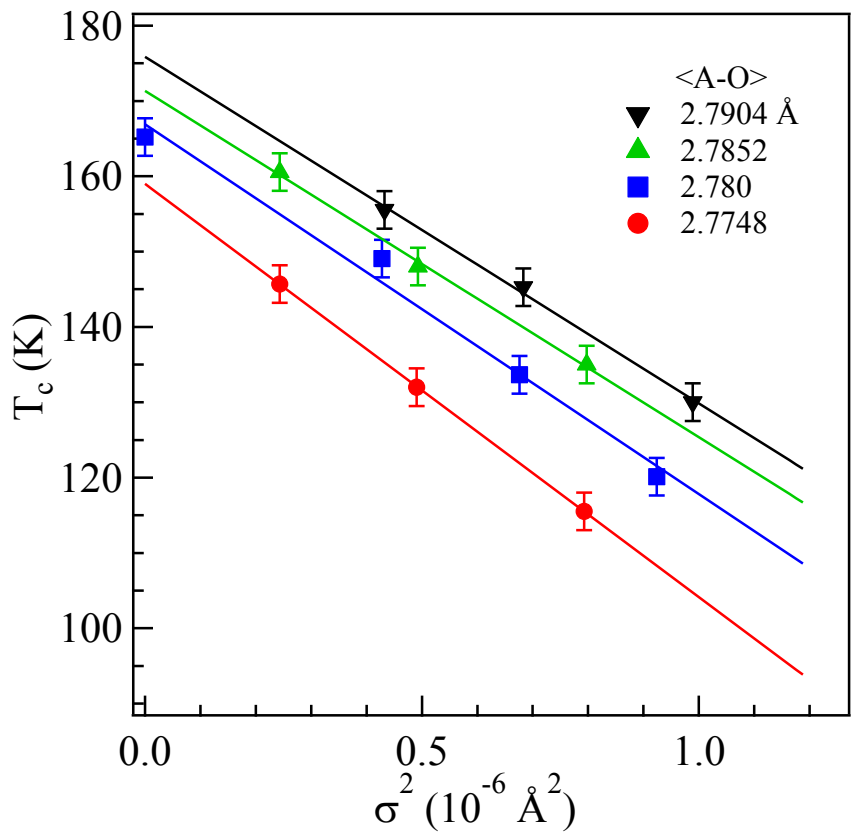

Fig.2 Ferromagnetic transition temperature $T_c$ versus the bond-length variance $\sigma^2$ for several series of perovskite ruthenates with different mean A-O bond length. Lines inside figure are results of linear fitting to the data.

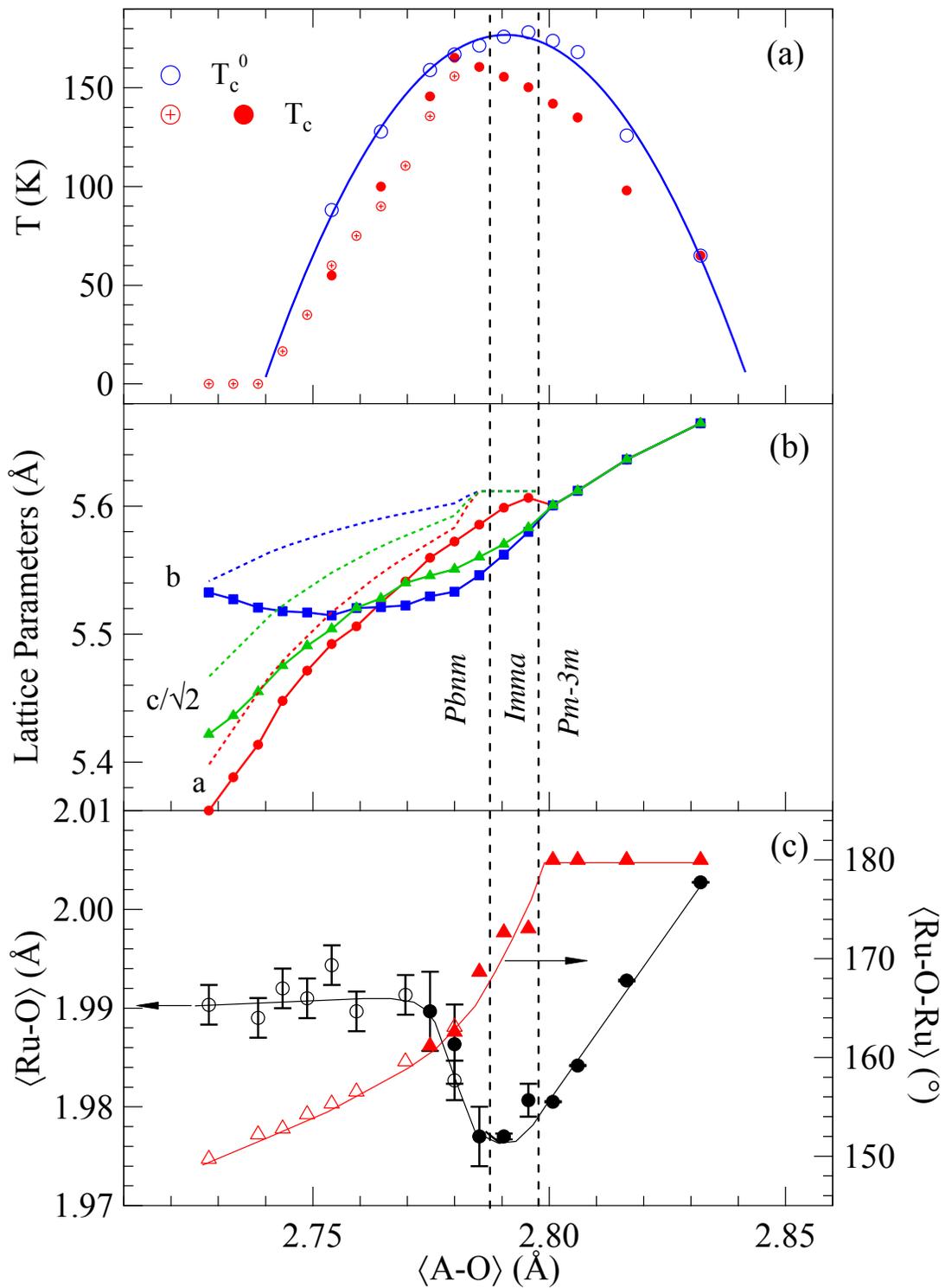

Fig.3 (a) The phase diagram of $T_c$ and $T_c^0$ (see the text for the definition) versus the mean A-O bond length $\langle A\text{-}O \rangle$. The solid symbol is for the samples synthesized under high pressure; the circle and plus symbol is for the samples synthesized under ambient pressure. (b) lattice parameters versus ⟨A-O⟩ on crossing three crystal structures ( we have converted lattice parameters from all other phases into the *Pbnm* structure for clarification). The dashed lines are lattice parameters calculated with SPuDS. (c) the Ru-O bond length and the Ru-O-Ru bond angle versus ⟨A-O⟩ in which data in open symbols are from ref.8 and those in solid symbols are from this work.